\documentclass[preprint,onecolumn,aps,nofootinbib]{revtex4}
\usepackage{graphicx}
\pdfoutput=1
\usepackage[colorlinks=true,linkcolor=blue,urlcolor=blue,filecolor=black,citecolor=red,pdfstartview=FitV,
pdftitle={},pdfsubject={},pdfkeywords={},pdfpagemode=None,bookmarksopen=true]{hyperref}
\usepackage{epsfig,ulem}
\usepackage{amsmath}
\usepackage{amsfonts}
\usepackage{amssymb}
\usepackage{color}%
\usepackage{dcolumn}
\providecommand{\U}[1]{\protect\rule{.1in}{.1in}}

\newcommand{\f}{\begin{equation}}
\newcommand{\ff}{\end{equation}}
\newcommand{\fa}{\begin{eqnarray}}
\newcommand{\ffa}{\end{eqnarray}}

\allowdisplaybreaks[4]
\begin{document}
\title{Holographic Butterfly Effect at Quantum Critical Points}
\author{Yi Ling $^{1,3,4}$}
\email{lingy@ihep.ac.cn}
\author{Peng Liu $^{1}$}
\email{liup51@ihep.ac.cn}
\author{Jian-Pin Wu $^{2,3}$}
\email{jianpinwu@mail.bnu.edu.cn}
\affiliation{$^1$ Institute of High Energy Physics, Chinese Academy of Sciences, Beijing 100049, China\ \\
$^2$ Institute of Gravitation and Cosmology, Department of
Physics,
School of Mathematics and Physics, Bohai University, Jinzhou 121013, China\ \\
$^3$ Shanghai Key Laboratory of High Temperature Superconductors,
Shanghai, 200444, China\ \\
$^4$ School of Physics, University of Chinese Academy of Sciences,
Beijing 100049, China}
\begin{abstract}
When the Lyapunov exponent $\lambda_L$ in a quantum chaotic system saturates the bound $\lambda_L\leqslant 2\pi k_BT$, it is proposed
that this system has a holographic dual described by a gravity theory. In particular, the butterfly effect as a prominent
phenomenon of chaos can ubiquitously exist in a black hole system characterized by a shockwave solution near the horizon. In this
paper we propose that the butterfly velocity can be used to diagnose quantum phase transition (QPT) in holographic theories.
We provide evidences for this proposal with an anisotropic holographic model exhibiting metal-insulator transitions (MIT), in which the
derivatives of the butterfly velocity with respect to system parameters characterizes quantum critical points (QCP) with local extremes
in zero temperature limit. We also point out that this proposal can be tested by experiments in the light of recent progress
on the measurement of out-of-time-order correlation function (OTOC).
\end{abstract}
\maketitle
\section{Introduction}
Quantum phase transition (QPT) is one of the essential and difficult topic in condensed matter theory (CMT).
It usually involves strong correlation physics where traditional treatments are inadequate.
Holographic duality has been proved a powerful tool to study strongly correlated system,
and has provided many novel insights into strongly correlated problems.
On the other hand, quantum chaos, also as known as butterfly effect, has been attracting unprecedented attention recently,
which set up a bridge among quantum theory, CMT and holographic gravity.
We shall address the connection between QPT and quantum chaos in holographic framework in this paper.

The butterfly effect states that an initially small perturbation becomes non-negligible at later time. The out-of-time-order correlation function (OTOC) in quantum systems can diagnose the butterfly effect by a sudden decay after the scrambling time $t_*$, which generically takes the following form,
\begin{equation}\label{otoc}
  F(t,\vec x)= \frac{\langle W^\dagger (t,\vec x) V^\dagger (0,0)
  W(t,\vec x) V(0,0) \rangle_\beta}{\langle W(t,\vec x) W(t,\vec x)
  \rangle_\beta \langle V(0,0)V(0,0)\rangle_\beta} = 1 - \alpha e^{\lambda_L\left( t - t_* - \frac{  |\vec x|}{v_B}\right)} + \cdots,
\end{equation}
where $W(t,\vec x)\equiv e^{iHt}W(0,\vec x)e^{-iHt}$, and $\langle\cdots\rangle_\beta$ represents the ensemble average at temperature $T=1/(k_B\beta)$.
$v_B$ is the butterfly velocity, $\lambda_L$ is the Lyapunov exponent and the scrambling time $t_*$ is the timescale when the commutator $[W(t,\vec x),V(0,0)]$ grows to
$\mathcal O(1)$. Physically, $F(t)$ describes the spread, or the scrambling of quantum information over the degrees of freedom across the system. Very importantly, as a characteristic
velocity of a chaotic quantum system, $v_B$ sets a bound on the speed of the information propagation \cite{Roberts:2016wdl}.

In holographic theories, the butterfly effect has extensively been studied in context \cite{Shenker:2013pqa,Shenker:2013yza,Roberts:2014isa,Roberts:2014ifa,
Shenker:2014cwa,Maldacena:2015waa,Polchinski:2015cea,Hosur:2015ylk,
Polchinski:2016xgd,Swingle:2016var,Blake:2016wvh,Blake:2016sud,Lucas:2016yfl}.
In the study of high energy scattering near horizon and information scrambling of black holes it is found that the butterfly effect ubiquitously exists and is signaled by a shockwave solution near the horizon \cite{Shenker:2013pqa,Shenker:2013yza,Shenker:2014cwa,Roberts:2016wdl,Blake:2016wvh} (see also \ref{Appendix-butterfly}). Especially, a bound on chaos has been proposed as
\begin{equation}\label{bound1}
\lambda_L \leqslant \frac{2\pi}{\beta},
\end{equation}
and the saturation of this bound has been suggested as a criterion on whether a many-body system has a holographic dual described by gravity theory \cite{Maldacena:2015waa}. One
remarkable example that saturates this bound is the Sachdev-Ye-Kitaev (SYK) model \cite{Kitaev:2014v1,Maldacena:2015waa}. Recently, the butterfly velocity $v_B$ has also been conjectured as the characteristic velocity that universally bounds the diffusion constants in incoherent metal \cite{Hartnoll:2014lpa,Blake:2016wvh,Blake:2016sud,Lucas:2016yfl}.

Since in holographic theories the bound in (\ref{bound1}) is always saturated,
we will focus on the behavior of the butterfly velocity close to quantum critical points
(QCP)\footnote{Previously, it was demonstrated in \cite{Shen:2016htm} that the Lyapunov exponent $\lambda_L$ may exhibit a peak near QCP in the Bose-Hubbard model.}.
The first signal to connect the butterfly velocity and QPT comes from the fact that both the butterfly velocity and the phase transition are controlled by IR degrees of freedom in chaotic quantum system \cite{Roberts:2016wdl,Donos:2012js}. This picture becomes more vivid in holographic scenario since IR degrees of freedom of the dual field theory is reflected by the near horizon data, and both $v_B$ and QPT depend solely on the near horizon data. In addition, the butterfly effect can be induced by any operator that affects the energy of the bulk theory \cite{Kitaev:2014v1,Roberts:2014ifa,Roberts:2016wdl}. Meanwhile, QPT is characterized by the degeneracy of ground states, which implies that the butterfly effect should be sensitive to QPT since they involve energy fluctuations. Therefore, it is highly possible that the butterfly effect can capture the QPT in holographic theories.

A heuristic argument about the relation between $v_B$ and QPT comes from the different behavior of the information propagation during the transition from a many-body localization (MBL) phase to a thermalized phase. A quantum system in MBL phase does not satisfy the Eigenstate Thermalization Hypothesis (ETH), and the quantum information propagates very slowly \cite{MBL1,Fan:2016ean}. In thermalized phase, however, the information propagates much faster. In other words, the speed of information propagation probably works as an indicator of a MBL phase transition. Notice that the butterfly velocity bounds the speed of the quantum information propagation across the chaotic system, it is reasonable to expect that the butterfly velocity may exhibit different behavior in distinct phases.

Inspired by above considerations, we propose that the butterfly velocity can characterize the QPT in generic holographic theories. We will present evidences for this proposal with a holographic model exhibiting MIT as an example of QPT, and demonstrate that the derivatives of the butterfly velocity with respect to system parameters do capture the QPT by showing local extremes near QCPs. Also, we point out the prospect of testing our proposal in laboratory.

\section{The butterfly effect and the quantum phase transition:}
In this section we demonstrate the relation between the MIT and the butterfly effects by numerical investigations on  holographic models.

In the context of gauge/gravity duality, holographic descriptions for the quantity in condensed matter physics can be computed in terms of the metric and other matter fields in the bulk.
On one hand, the holographic description of MIT has been studied in \cite{Donos:2012js,Donos:2013eha,Ling:2015dma,Ling:2016wyr}. Usually, the transition is induced by relevant deformations to near horizon geometry, in which the lattice structure plays a key role. In this paper we consider the holographic Q-lattice model exhibiting MIT, which is presented in \ref{Appendix-MIT1} (for more details, refer to \cite{Donos:2013eha,Ling:2015dma}).
On the other hand, the butterfly effect in black holes has been investigated in \cite{Dray:1984ha,Kiem:1995iy,Roberts:2016wdl,Blake:2016wvh,Cai:2017ihd}, and the butterfly velocity can be extracted from shockwave solutions to the perturbation equations of gravity. Since the bulk geometry we consider here is anisotropic, we present a detailed derivation for the corresponding $v_B$ in section \ref{Appendix-butterfly}.

Next, we introduce the holographic Q-lattice model and the {\it anisotropic} holographic butterfly effects. After that, we explicitly provide the numerical evidences for our proposal. Moreover, we also study the anisotropy of the butterfly velocity and its effects on the connection between QPT and butterfly effects.

\subsection{Holographic Q-lattice model}\label{Appendix-MIT1}
The Lagrangian of the holographic Q-lattice model reads as
\cite{Donos:2013eha,Donos:2014uba,Ling:2015dma},
\begin{equation}\label{qlaact}
\mathcal{L}=R+6-\frac{1}{4}F^2-|\nabla \Phi|^2-m^2|\Phi|^2,
\end{equation}
where $F=dA$ is the field strength of the Maxwell field and $\Phi$
is the complex scalar field simulating the Q-lattice
structure. Note that we have set the AdS radius $L=1$, and we adopt the natural system of units where $c, k_B, h$ are set to $1$.
The equations of motion corresponding to \eqref{qlaact} read as
\begin{eqnarray}
   R_{ab} + \frac{g_{ab}}{2} \left( 6 - m^2 |\Phi|^2 \right) -\partial_{(a} \Phi\partial_{b)}\Phi^* -\frac{1}{8}\left( 4F^2_{ab} -g_{ab} F^2  \right) &=& 0,  \\
   \nabla_a\nabla^a \Phi-m^2 \Phi &=& 0, \\
   \nabla_a F^{ab} &=& 0.
\end{eqnarray}

The ansatz for a black brane solution with lattice structure only along $x$
direction is presented as
\begin{eqnarray}
&&
d{s^2} =\frac{{{1}}}{{{z^2}}}\left( { - fSd{t^2} + \frac{{d{z^2}}}{{fS}} + {\hat V_x}d{x^2} + {\hat V_y}d{y^2}} \right),
\nonumber
\\
&&
A_t=\mu(1-z)a,\,\,\,\,\,\,\Phi = e^{i \tilde{k} x}z^{3-\Delta}\phi,
\label{metric-q}
\end{eqnarray}
where $f(z)\equiv (1-z)(1+z+z^2-\mu^2z^3/4)$ and
$\Delta=3/2\pm(9/4+m^2)^{1/2}$. $S,\hat V_x,\hat V_y,a$ and $\phi$
are functions of the radial coordinate $z$ only and $\mu$ corresponds to the
chemical potential of the dual field theory by setting the boundary condition $a(0)=1$. Black brane solutions
are obtained by numerically solving the Einstein equations as well
as other equations of motion for matter fields.
System \eqref{metric-q} is invariant under scaling $\{ z,t,x,y \} \to \alpha\{ z,t,x,y \}, \{\mu,k\}\to \{\mu,k\}/\alpha,
 \{ g_{tt}, g_{zz},g_{xx},g_{yy}\} \to \{ g_{tt}, g_{zz},g_{xx},g_{yy}\}/\alpha^2$.
We only focus on the scaling dimensionless quantities by taking the chemical potential $\mu$ as scaling unit,
which means that we are effectively working with grand ensemble description.
Each solution is specified by three dimensionless parameters, namely the
temperature $\tilde T/\mu$ with $\tilde T=(12-\mu^2)S(1)/16\pi$,
lattice amplitude $\tilde\lambda/\mu^{3-\Delta}$ with
$\tilde\lambda\equiv\phi(0)$, and lattice wave number $\tilde
k/\mu$, which are abbreviated as $\{T,\lambda,k\}$ in this paper.
The metric has an event horizon at $z=1$ and the spacetime
boundary locates at $z=0$. Throughout this paper, we set
$m^{2}=-2$ such that the scaling dimension of $\Phi$ is $\Delta =2$.
We would like to point out that for other values of $m^2$, qualitatively similar phenomena
will be obtained. Moreover, we have also examined the case $\Delta =1$ for $m^2=-2$, and similar
results to the case $\Delta=2$ are obtained as well.

\begin{figure}
  \centering
    \includegraphics[width=0.6\textwidth]{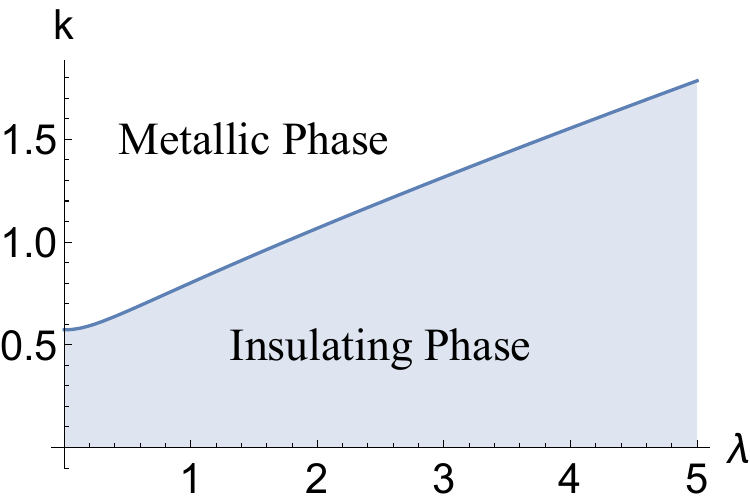}
    \caption{MIT phase diagram at T = 0.001 \cite{Ling:2015dma}.}
    \label{phasediagram1}
\end{figure}

The occurrence of MIT in this model has been discussed in
\cite{Donos:2013eha} and an explicit phase diagram over $(\lambda,
k)$ plane (Fig. \ref{phasediagram1}) has been presented in \cite{Ling:2015dma}, where the
temperature is fixed at $T\sim 10^{-3}$, but further decreasing
the temperature will not induce significant modifications to the phase diagram.
From Fig. \ref{phasediagram1} it is seen that increasing $\lambda$ at certain value of $k$ will
drive the system from metallic phase into insulating phase, which is consistent with
the interpretation of $\lambda$ as the lattice strength.

At finite but extremely low temperature, we distinguish the metallic
phase and the insulating phase by the different temperature
dependence of DC conductivity. Specifically, the metallic phase is
defined by $\partial_T \sigma_{DC}(T)<0$ while insulating phase
$\partial_T \sigma_{DC}(T)>0$, therefore the surface $\partial_T
\sigma_{DC}(T)=0$ separating the insulating phase and the metallic
phase is the critical surface. This criterion has also been widely
adopted in holographic literature
\cite{Baggioli:2014roa,Ling:2015dma,Ling:2016wyr}.

The expressions of DC conductivity $\sigma_{DC}$ along x-direction
for model \ref{Appendix-MIT1} can be calculated from
\begin{equation}\label{dc}
  {\sigma_{DC}=\left.
\left( {\sqrt {\frac{{{\hat V_y}}}{{{\hat V_x}}}}  + \frac{{{\mu
^2}{a^2}\sqrt {{\hat V_x}{\hat V_y}} }}{{{k^2}{\phi ^2}}}} \right)\right|_{z
= 1},}
\end{equation}
which are determined by the horizon data \cite{Donos:2014uba,Ling:2015dma,Ling:2016wyr}.
Alternatively, one may compute the $\sigma_{DC}$ as the zero frequency limit of the optical
conductivity $\lim_{\omega \to 0}\sigma(\omega)$ by introducing the following consistent
time-dependent perturbation,
\begin{equation}\label{a1-qla-perturbation}
  \delta A_x = a_x (z)e^{-i\omega t}, \; \delta g_{tx}=h_{tx}(z)e^{-i\omega t},\;\delta \Phi=ie^{ikx}z^{3-\Delta}\varphi(z)e^{-i\omega t}.
\end{equation}
After numerically solving the perturbation equation of motions on a numerical background
solution to \eqref{qlaact}, the optical conductivity can be obtained as
\begin{equation}\label{a5-cond}
\sigma(\omega)=\left.\frac{\partial_z a_x(z)}{i\omega a_x (z)}\right|_{z=0}.
\end{equation}

\subsection{The anisotropic butterfly velocity}\label{Appendix-butterfly}
In this section we demonstrate the derivation of butterfly
velocity $v_B$ in anisotropic background, which can be extracted
from the shockwave solution near the horizon
\cite{Dray:1984ha,Kiem:1995iy,Blake:2016wvh,Roberts:2016wdl}. For
this purpose, it is more convenient to work in $r$-coordinate with
$r\equiv r_0/z$, where $r_0$ is the location of horizon. The
generic spatially anisotropic metric of a 4-dimensional spacetime
can be written as
\begin{equation}\label{metric}
  ds^2 = - U(r) dt^2 + \frac{dr^2}{U(r)} + V_x(r)dx^2 + V_y(r)dy^2.
\end{equation}
In Kruskal coordinate (\ref{metric}) is written as
\begin{equation}\label{krumetric}
  ds^2 = \mathcal{U}(uv)dudv + V_x (uv) dx^2 + V_y(uv)dy^2,
\end{equation}
where $uv=-e^{U'(r_0)r_*(r)},u/v=-e^{-U'(r_0)t}$, with $r_*$ being
the tortoise coordinate defined by $dr_* = dr/U(r)$. In addition,
$\mathcal
U(uv)=\frac{4U(r)}{uvU'(r_0)^2},\,V_{x,y}(uv)=V_{x,y}(r)$. Note
that, in this coordinate the horizon is at $u=v=0$.

The shockwave geometry is induced by a freely falling
particle on the AdS boundary at $t_i$ in the past and at $x=y=0$.
This particle is exponentially accelerated in Kruskal
coordinate and generates the following energy distribution
at $u=0$,
\begin{equation}\label{tab}
  \delta T_{uu} \sim E_0 e^{\frac{2\pi}{\beta} t_i} \delta(u)\delta(x,y),
\end{equation}
where $E_0$ is the initial asymptotic energy of the particle.
After the scrambling time $t_* \sim \beta \log N^2$ an initially small perturbation
becomes significant and back-react to the geometry by a shockwave localized at
the horizon \cite{Sekino:2008he},
\begin{equation}\label{petmetric}
\begin{aligned}
  ds^2 = & V_x(uv) dx^2 + V_y(uv) dy^2  + \mathcal U(uv)dudv\\
&- \mathcal U(uv)\delta(u)h(x,y)du^2.
\end{aligned}
\end{equation}
By a convenient redefinition $\tilde y \equiv  y\sqrt{\frac{V_x(0)}{V_y(0)}}$ the
resultant Einstein equation can be written as a Poisson equation,
\begin{equation}\label{anieq}
  \left( \partial^2_x + \partial^2_{\tilde y} - m^2 \right) h(x,\tilde y) \sim
  \frac{16\pi G_N V_x(0)}{\mathcal U(0)}E_0 e^{\frac{2\pi}{\beta} t_i}\delta(x,\tilde y),
\end{equation}
with $m^2$ given by
\begin{equation}\label{newm}
m^2 = \left.\frac{4}{\mathcal U(uv)}\left( V_x'(uv) + \frac{V_x(uv)V_{\tilde y}'(uv)}{V_{\tilde y}(uv)} \right)\right|_{u=0}.
\end{equation}
At long distance $|\vec x| \equiv \sqrt{x^2 + \tilde y^2} \geqslant m^{-1}$, the solution reads as
\begin{equation}\label{sol1}
  h(x,\tilde y) \sim \frac{E_0 e^{\frac{2\pi}{\beta}(t_i - t_*)-m|\vec x|}}{|\vec x|^{1/2}}.
\end{equation}
From (\ref{sol1}) we read off the Lyapunov exponent $\lambda_L$ and the butterfly velocity $v_B$,
\begin{equation}\label{expo}
  \lambda_L =\frac{2\pi}{\beta}, \quad v_B = \frac{2\pi}{\beta m}
\end{equation}
The Lyapunov exponent saturates the chaos bound as expected.
Rewriting $m$ in coordinates
$(r,t,x,\tilde y)$ we find
\begin{equation}\label{vbe}
  v_B= \sqrt{\frac{\pi T V_{\tilde y}(r_0)}{V_x'(r_0) V_{\tilde y}(r_0)+ V_x(r_0)V_{\tilde y}'(r_0)}}.
\end{equation}
When recovered to $(x,y)$ coordinate system, the butterfly
velocity $\mathfrak{v}_B$ is anisotropic. Specifically, in
direction with polar angle $\theta$,
\begin{equation}\label{vbe2}
\mathfrak{v}_B(\theta)  = v_B \sqrt{\frac{\sec ^2(\theta )
V_x\left(r_0\right)}{V_x\left(r_0\right)+\tan ^2(\theta )
V_y\left(r_0\right)}}.
\end{equation}

\subsection{Evidences from holographic theories}
In this subsection we explicitly study the relation between the QPT and the butterfly velocity in Q-lattice model. We approach the QPT by studying the phase transitions in zero temperature limit. Therefore, our main task is to investigate the butterfly velocity on background solutions specified by $(\lambda,k)$, which correspond to lattice strength and wave number, respectively. For simplicity, we focus on $\lambda=2$ and study the behavior of the butterfly velocity along $x$-direction, {\it i.e.}, $v_B$ over $k$ in low temperature region \footnote{Very similar phenomena can be obtained when varying $\lambda$ with fixed $k$.}. The anisotropy of the butterfly velocity will be addressed in the next subsection.

\begin{figure}
\begin{center}
  \includegraphics[width=0.8\textwidth]{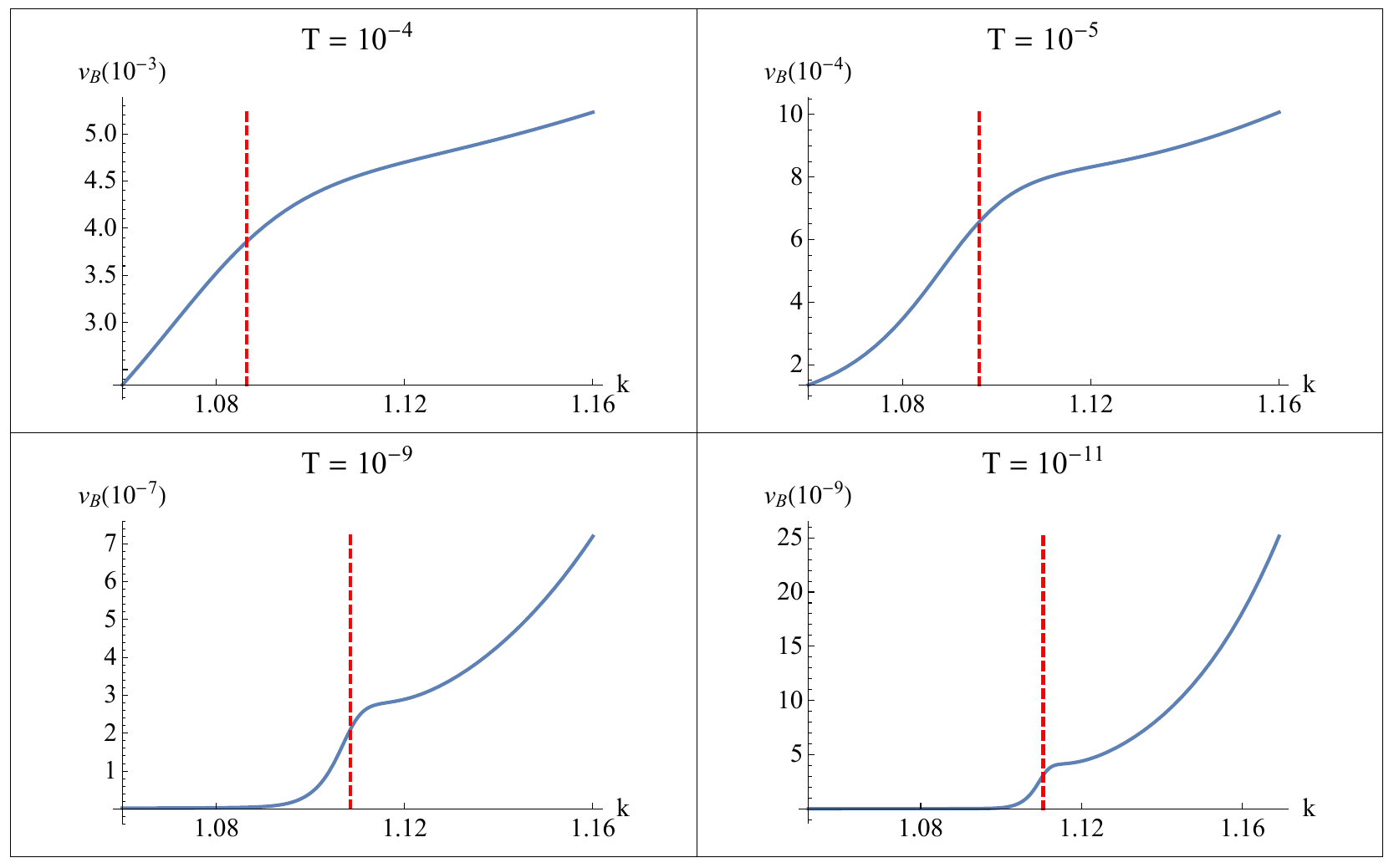}
  \caption{$v_B$ v.s. $k$ at different low temperatures
  $T=10^{-4},10^{-5},10^{-9},10^{-11}$ respectively.
  In each plot the dotted line in red represents the location of QCP,
 separating the insulating phase (left side) and metallic phase (right side).}
  \label{vbdt}
\end{center}
\end{figure}

\begin{figure}
  \centering
  \includegraphics[width=0.45 \textwidth]{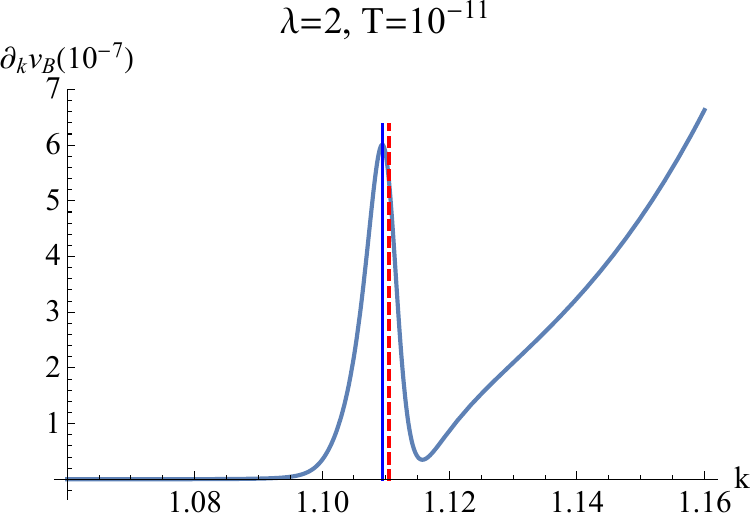}
  \includegraphics[width=0.45 \textwidth]{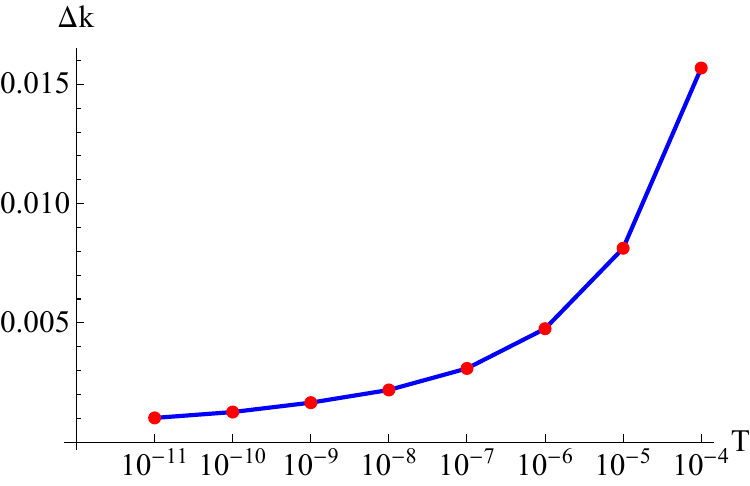}
  \caption{The left plot is $\partial_k v_B \,v.s.\, k$ at $T=10^{-11}$,
  in which the red vertical line represents the position of the critical point
  while the blue line denotes the position of the local maximum of $\partial_k v_B$.
  The right plot is for the temperature dependence of $\Delta k$.}
  \label{temdep11}
\end{figure}

First, we plot $v_B\,v.s.\, k$ at low temperatures in Fig. \ref{vbdt}. It is seen that $v_B$ becomes larger when the system transits from insulating phase to metallic phase. In particular, $v_B$ in insulating phases is always several orders of magnitude smaller than that in metallic phases. Therefore, it can be expected that the critical points can be captured by local extremes of derivatives of $v_B$ with respect to $k$. We confirm this expectation in the left plot of Fig. \ref{temdep11}. It is obvious that the location of the local maxima of $\partial_k v_B$ is always close to critical points\footnote{Singular behaviors happen at the quantum critical points at absolute zero temperature. In this paper, we investigate the quantum critical phenomena by working at ultra low temperature, which is still finite. Our system is regular at any finite temperature, and hence the $v_B$ is smooth function of system parameter $\lambda, k$.}.

Moreover, we demonstrate the phenomenon that local extremes of $\partial_k v_B$ captures QPT is robust in zero temperature limit. Specifically, we show $\Delta k$, denoting the difference between the locations of the critical point and local maxima of $\partial_k v_B$, as the function of the temperature in the right
plot of Fig. \ref{temdep11}, and find that $\Delta k$ continuously decreases with temperature. Therefore we arrive at the conclusion that in Q-lattice
model \ref{Appendix-MIT1} the local extreme of $\partial_k v_B$ can be used to characterize the QPT in zero temperature limit.

\begin{figure}
  \centering
  \includegraphics[width=0.6 \textwidth]{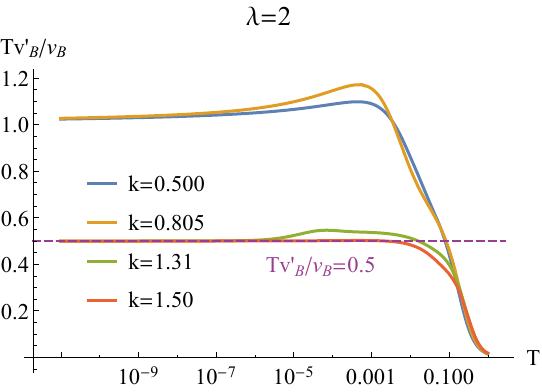}
  \caption{$Tv'_B/v_B \,v.s.\, T$ for different phases ($k=0.500,\,0.805$ corresponds
  to metallic phases and $k=1.31,\,1.50$ corresponds to insulating phases).
  The purple dashed line points to $Tv'_B/v_B=0.5$.}
  \label{sca1}
\end{figure}

Inspired by the fact that in zero temperature limit $v_B$ tends to vanish for
both metallic phases and insulating phases, we intend to understand the above phenomena by studying the scaling of $v_B$ with temperature $v_B \sim T^\alpha$
in both metallic and insulating phases. Fig. \ref{sca1} demonstrates $Tv'_B/v_B$ as a function of $T$,
which captures the exponent $\alpha$ in different phases. One
finds $\alpha =1/2$ for metallic phases in low temperature region.
This originates from the fact that metallic phases in Q-lattice model \ref{Appendix-MIT1}
always correspond to the well-known AdS$_2 \times \mathbb R^2$ IR geometry,
on which $v_B \sim T^{1/2}$ can be deduced \cite{Blake:2016sud}.
While for insulating phases, $Tv'_B/v_B$ tends to converge to
a fixed value close to $1$ down to ultra low temperature
$T=10^{-11}$, which implies that the insulating phases for model \ref{Appendix-MIT1}
may correspond to a single IR geometry different from AdS$_2\times \mathbb{R}^2$ \footnote{However,
we would like to point out that the exact IR fixed point of Q-lattice model is unknown so far \cite{Donos:2013eha}.}.
Therefore we conclude that $v_B$ scaling distinctly with temperature in metallic
phases and insulating phases, are responsible for the rapid change of $v_B$
observed in Fig. \ref{vbdt}, as well as the local extremes of $\partial_k v_B$
near QCPs observed in Fig. \ref{temdep11}.

The above understanding is also applicable for some other holographic MIT model. To this end, we demonstrate another holographic model \cite{Donos:2014uba}, in which MIT is also achieved when varying the system parameters in the region $-1/3 < \gamma \leqslant -1/12$, where $\gamma$ is the parameter of the action. Like model \ref{Appendix-MIT1}, we obtain $v_B \sim T^{1/2}$ in metallic phases again, due to the AdS$_2 \times \mathbb R^2$ IR geometry. While for insulating phases we find $\alpha = {\frac{2 \gamma ^2+7 \gamma +21}{2 \gamma ^2+4 \gamma +18}}$, which reduces to $\frac{85}{76} < \alpha \leqslant \frac{1471}{1273}$ in terms of the range of $\gamma$. Therefore, the QPT of model in \cite{Donos:2014uba} can also be characterized by derivatives of $v_B$ with respect to system parameters.

\subsection{Anisotropy of the butterfly velocity in quantum critical region}
In previous subsection we disclose the
connection between QPT and the butterfly velocity  along $x$-direction for simplicity. Here
we study the anisotropy of the butterfly velocity $\mathfrak v_B(\theta)$ in quantum critical region.
Since the period of $\mathfrak{v}_B (\theta)$ is $\pi$
and $\mathfrak{v}_B (\pi/2-\theta) = \mathfrak{v}_B (\pi/2+\theta)$ (see Eq. \eqref{vbe2}),
we shall only focus on the angle range $\theta \in [0,\pi/2]$ in what follows.

\begin{figure}
  \centering
  \includegraphics[width=0.6 \textwidth]{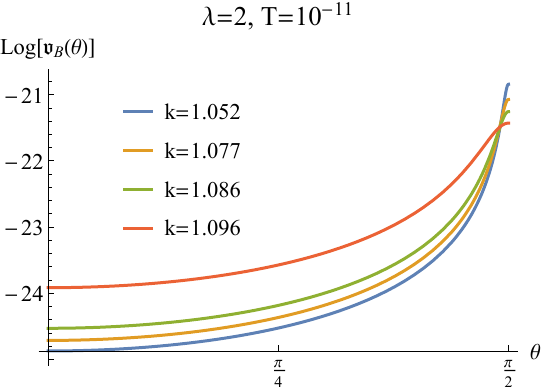}
  \caption{$\mathfrak{v}_B (\theta)\,v.s.\,\theta$ at $\lambda=2, T = 10^{-11}$ with $k$ specified by the plot legend.}
  \label{angdep}
\end{figure}

First, we demonstrate $\mathfrak v_B (\theta)\,v.s.\,\theta$ at $T=10^{-11},\,\lambda =2$ in Fig. \ref{angdep}, from which one can see that
$\frak v_B(\theta)$ monotonically increases. In other word, $v_B$ along the latticed $x$-direction $(\theta =0)$ is always
smaller that along $y$-direction ($\theta=\pi/2$). Therefore, the lattice {\bf suppresses} the butterfly velocity.
This phenomenon originates from the fact that  in Q-lattice model $V_x(1)>V_y(1)$, which has been verified in our numerics,
consequently $\mathfrak{v}_B(\theta) = v_B [{\cos^2(\theta)+\sin^2(\theta)V_y(1)/V_x(1)}]^{-1/2}$ (see \eqref{vbe2})
monotonically decreases with $\theta$.

\begin{figure}
  \centering
  \includegraphics[width=0.45 \textwidth]{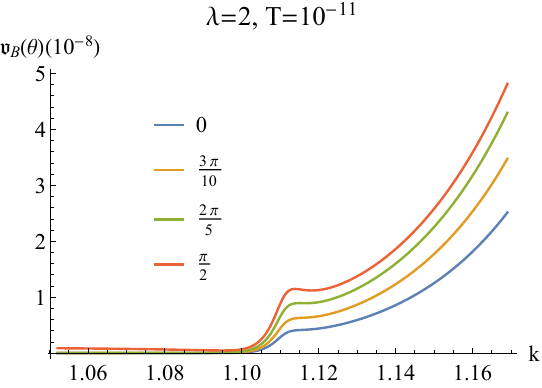}
  \includegraphics[width=0.45\textwidth]{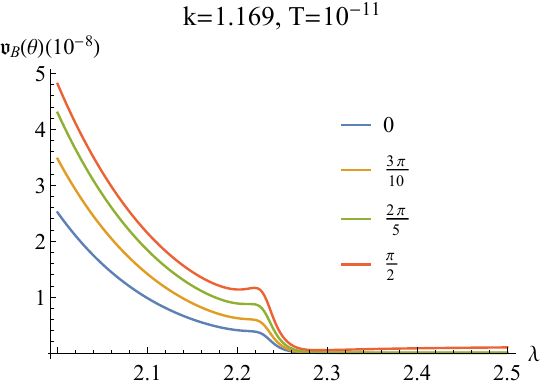}
  \caption{Left plot: $\mathfrak{v}_B (\theta)$ at $\lambda=2, \, T = 10^{-11}$. Right Plot: $\mathfrak{v}_B (\theta)$ at $k=1.169, \, T = 10^{-11}$. Four curves in each plot corresponds to   different angle $\theta$ specified by the plot legend. }
  \label{difflk2}
\end{figure}

Next, we show $\mathfrak v_B(\theta)\,v.s.\,k$ at $\lambda=2$ and $\mathfrak v_B(\theta)\,v.s.\,\lambda$ at $k=1.169$ in Fig. \ref{difflk2}.
We can see that when $\theta$ is small, $\mathfrak v_B(\theta)$ monotonically increases with $k$ and $\mathfrak v_B(\theta)$
monotonically decreases with $\lambda$. However, while $\theta$ is relatively large the monotonicity changes.

\begin{figure}
  \centering
  \includegraphics[width=0.6 \textwidth]{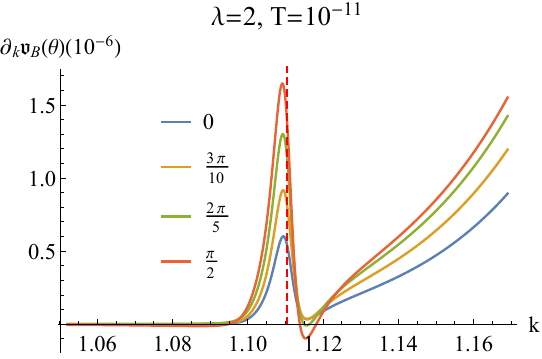}
  \caption{$\partial_k \mathfrak{v}_B(\theta)$ at different angles specified by the plot legends where the red vertical line represents the position of the critical point.}
  \label{pdkvbangle}
\end{figure}

At last, we point out that the $\mathfrak v_B (\theta)$ can diagnose the QPT in any direction.
As we can see from Fig. \ref{pdkvbangle} that $\partial_k \mathfrak{v}_B(\theta)$ reaches local maximums near the QCP in any direction.
This phenomenon reflects the fact that the different IR fixed point leads to distinct behavior of $\mathfrak v_B (\theta)$.

\section{Discussion}
In model \ref{Appendix-MIT1} we have demonstrated that the derivatives of
$\mathfrak{v}_B (\theta)$ with respect to system parameters diagnoses the QCP with
local extremes in zero temperature limit.
The underlying reason is that IR fixed points of the metallic phases and
insulating phases are distinct. A direct connection between derivatives of $v_B$ with respect
to system parameters and another criterion, $\partial_T\sigma_{DC}$, may be disclosed
by analytical analysis.
As an extension we believe that the scenario of $v_B$
characterizing QPT is applicable to other holographic models with
MIT (for instance the isotropic lattice model, the lattice with
helical symmetry or massive gravity), and also to those
exhibiting other sorts of QPT. Of course in these
circumstances, other than local extremes of $\partial_k v_B$
originating from the distinct $v_B$ scaling with temperature in
different phases, the characterizing style of $v_B$ can be more diverse.

Although our evidences come from holographic theories that always
saturate the chaos bound, our proposal may also apply for chaotic
quantum system that does not saturate the bound. A direct argument
is that IR dependence of $v_B$ and QPT does not require a
holographic theory.

More importantly, our proposal can be tested by experiments in
light of recent progress on the measurement of OTOC.
Experimentally, the butterfly velocity $v_B$, and its relation to
QPT, can be studied by measuring the OTOC of a QPT system.
Recently, new protocols and methods, that are versatile to
simulate diverse many-body systems and achievable with
state-of-the-art technology, have been proposed to measure the
OTOC \cite{Swingle:2016var,Yao:2016ayk}. Furthermore, experimental
measurements of the OTOC have also been implemented
\cite{Garttner:2016mqj,zhai:2016exp}. All these progresses provide
test beds for our proposal.

Our work has offered an information-theoretic diagnose of the QPT.
The distinct behavior of information propagation in a
quantum many-body system may signalize different phases.
This phenomenon indicates that the information-theoretic property of a
chaotic many-body system can work as a novel tool to study
QPT. It can be expected that more insights into QPT will be gained
from the quantum information theory.

{\it Acknowledgements -} We are very grateful to Wei-Jia Li, Ya-Wen Sun,
Meng-he Wu, Zhuo-yu Xian, Yi-kang Xiao, Xiao-Xiong Zeng and Xiang-rong Zheng for helpful
discussion. This work is supported by the Natural Science
Foundation of China under Grant Nos.11275208, 11305018,
11575195 and 11775036, and by the grant (No. 14DZ2260700) from the Opening
Project of Shanghai Key Laboratory of High Temperature
Superconductors. Y.L. also acknowledges the support from Jiangxi
young scientists (JingGang Star) program and 555 talent project of
Jiangxi Province. J. P. Wu is also supported by Natural Science Foundation of Liaoning Province under
Grant Nos.201602013.

\end{document}